\documentclass[%
 reprint,
 superscriptaddress,
 amsmath,amssymb,
 aps,
]{revtex4-2}

\usepackage{graphicx}
\usepackage{dcolumn}
\usepackage{bm}

\usepackage{textcomp}
\usepackage{upgreek}

\begin{document}

\preprint{}

\title{Slit-loaded coplanar waveguide for color-center spin qubits}

\author{Haruko Toyama}
 \email{haruko.toyama.zd@mosk.tytlabs.co.jp}
\author{Kosuke Tahara}
\author{Taro Ikeda}
\author{Hiroya Tanaka}
\author{Atsushi Miura}
\author{Shin-ichi Tamura}

\affiliation{%
 Toyota Central R\&D Labs., Inc., Nagakute, Aichi 480-1192, Japan
}%

\author{Maria Emma Villamin}
\author{Toshinori Numata}
\author{Naotaka Iwata}

\affiliation{%
 Research Center for Smart Energy Technology, Toyota Technological Institute, 2-12-1, Hisakata, Tempaku-ku, Nagoya, Aichi 468-8511, Japan
}%

\author{Yuichi Yamazaki}
\affiliation{%
 National Institutes for Quantum Science and Technology (QST), 1233 Watanuki, Takasaki, Gunma 370-1292, Japan
}%

\author{Takeshi Ohshima}
\affiliation{%
 National Institutes for Quantum Science and Technology (QST), 1233 Watanuki, Takasaki, Gunma 370-1292, Japan
}%
\affiliation{%
 Department of Materials Science, Tohoku University, Aoba, Sendai, Miyagi 980-8579, Japan
}%

\author{Katsuhiro Kutsuki}
\author{Hideo Iizuka}
 \thanks{hiizuka@mosk.tytlabs.co.jp}

\affiliation{%
 Toyota Central R\&D Labs., Inc., Nagakute, Aichi 480-1192, Japan
}%


\begin{abstract}
The spin qubits of color centers are extensively investigated for quantum sensing, communication, and information processing, with their states generally controlled using lasers and microwaves. 
However, it is challenging to effectively irradiate both lasers and microwaves onto color centers using small footprint microwave waveguides or antennas that are compatible with semiconductor devices. 
We experimentally show that by introducing a compact coplanar waveguide with a thin slit in its signal line, effective irradiation of both lasers and microwaves is enabled, allowing spin-state control of color centers created around the slit. 
Microwave magnetic fields parallel to the surface, intrinsically generated by a standard coplanar waveguide, persist even after loading the slit, which is necessary to control the color centers whose spin quantization axes are oriented perpendicular to the surface, while laser light for the initialization and readout of spin states can access the color centers through the slit. 
Continuous and pulsed optically detected magnetic resonance measurements are performed for the silicon vacancies ($\mathrm{V_{Si}}$) in silicon carbide 4H-SiC(0001). 
Experimental results indicate that the spin states of $\mathrm{V_{Si}}$ are effectively controlled by the microwave magnetic fields parallel to the surface, which agrees with numerical results from electromagnetic field simulations. 
Our small footprint waveguide is suitable for integrating color-center-based quantum sensors into semiconductor electronic devices and other small-scale systems. 
\end{abstract}

\maketitle


\section{\label{sec:level1}Introduction}

Color centers in solids acting as spin qubits under ambient conditions have attracted considerable attention for quantum technologies such as quantum sensing with high spatial resolution at the nanoscale and high sensitivity, and quantum communication using them as single-photon sources \cite{ref01,ref02,ref03}. 
Among color centers, negatively charged nitrogen-vacancy ($\mathrm{NV^{-}}$) center in diamond has been extensively investigated due to its long coherence time and capability of optical spin-state initialization and readout even at room temperature \cite{ref04,ref05,ref06}. 
Consequently, numerous demonstrations of sensing and imaging of magnetic fields \cite{ref07,ref08,ref09,ref10}, electric fields \cite{ref11,ref12,ref13}, pressures \cite{ref14}, and temperatures \cite{ref15,ref16} based on diamond NV centers have been reported. 
Additionally, several color centers in silicon carbide (SiC) are also promising candidates with optically controllable and addressable spin states \cite{ref17,ref18,ref19}, among which the negatively charged silicon vacancy ($\mathrm{V_{Si}}$) center at the cubic site in 4H-SiC (also known as the V2 center), has been extensively investigated \cite{ref20,ref21,ref22,ref23,ref24}. 
Hereafter, we will refer to the V2 center as $\mathrm{V_{Si}}$. 
Unlike diamond, SiC has been commercialized for power electronics devices, ensuring a stable supply of materials and well-established device process technology, both of which are advantageous for industrial applications.

Microwave engineering is crucial for enhancing the performance of spin-based quantum technology. 
A microwave system for color-center-based sensors basically needs to efficiently generate microwave fields oriented in the required direction while ensuring a light path for the initialization and readout of spin states. 
When focusing on semiconductor device integration, a compact system size and wide frequency range are additionally required. 
However, a microwave mechanism achieving these aforementioned features has not been reported.

For example, a thin wire with a diameter of tens of micrometers is a simple method to generate relatively strong microwave magnetic fields typically localized within several tens of micrometers from the wire \cite{ref08}. 
Yet, the manual attachment of the wire to materials hinders reproducibility and packaged system fabrication.
Microwave resonators are commonly used for widefield imaging applications due to their high spatial uniformity over $\mathrm{mm^2}$ areas \cite{ref25,ref26,ref27,ref28,ref29,ref30,ref31}. 
Some resonator designs focus on expanding \cite{ref26,ref30} or tuning \cite{ref27} the resonant frequency range, enabling extensive magnetic field and temperature measurements, polarization control for selective transitions \cite{ref27}, and enhanced noise robustness \cite{ref28}. 
Unlike wires, resonators offer the advantage of direct integration into host materials of color centers, providing greater convenience for device incorporation. 
However, since the resonator size is inherently related to the resonant frequency of generated microwaves, both miniaturization and wide frequency range are difficult to achieve. 
For instance, the 10 to 1000 MHz resonant frequencies of diamond NV and 4H-SiC $\mathrm{V_{Si}}$ correspond to microwave wavelengths of 10 to 1000 mm in solids, leading to typical resonators with sizes exceeding 10 mm.

In contrast, planar waveguides with signal and ground lines on the same plane, such as coplanar waveguides, provide a wide frequency range and can be designed with narrower transmission lines \cite{ref32,ref33,ref34}. 
Furthermore, some resonators integrated into these waveguides \cite{ref35,ref36,ref37,ref38,ref39} also achieve millimeter-scale compactness. 
However, unlike three-dimensional antennas \cite{ref40}, microwave magnetic fields induced by most planar waveguides are approximately oriented perpendicular to the surface at the location where color centers, which demand optical measurements, are created. 
This imposes a limitation on spin control of color centers because the efficiency of the spin manipulation is maximized when the microwave magnetic fields are perpendicular to the spin quantization axis and is zero when they are parallel to it. 
Consequently, a compact planar waveguide capable of generating fields parallel to the surface is required for wider applicability, particularly for color centers with spin quantization axes perpendicular to the surface, such as the $\mathrm{V_{Si}}$ centers in 4H-SiC(0001) and NV centers in diamond(111).

In this work, we experimentally demonstrate a coplanar waveguide with a thin slit in its signal line, enabling the simultaneous exposure of color centers to both laser light and microwaves for controlling their spin states. 
Simulation results show that an in-plane microwave magnetic field component $B_{\mathrm{AC},x}$ is generated within the slit. 
Since this slit allows laser light to reach color centers within it, the spin states of color centers with spin quantization axes oriented perpendicular to the solid surface can be controlled. 
We perform continuous-wave (CW) optically detected magnetic resonance (ODMR) and Rabi oscillation measurements for the $\mathrm{V_{Si}}$ (V2) centers in 4H-SiC(0001) using this waveguide, which features a 40 $\upmu$m wide slit within its 100 $\upmu$m wide signal line. 
The applied $B_{\mathrm{AC},x}$ strengths, estimated from the measured Rabi frequency, show reasonable agreement with simulation results, including the spatial dependence within the slit. 
In addition, we discuss the depth dependence of $B_{\mathrm{AC},x}$ and show that the depth at which $B_{\mathrm{AC},x}$ is maximized can be tuned by adjusting the widths of the slit and the signal line. 
Our slit-loaded coplanar waveguide will advance the development of the color-center-based semiconductor electronic devices.

\section{\label{sec:level1}Waveguide design}

\begin{figure}[b]
\includegraphics{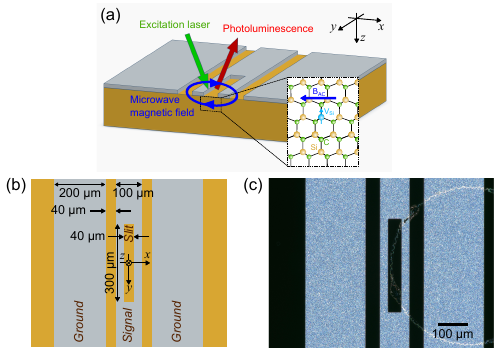}
\caption{\label{fig:fig1}
  (a) Illustration of our slit-loaded coplanar waveguide. The gray and yellow regions represent the parts made of metal (Al) and dielectric (4H-SiC), respectively. The signal line has a thin slit, which enables excitation laser irradiation (green arrow) and PL extraction (red arrow). Even with a slit, the microwave magnetic fields ($\boldsymbol{B}_{\mathrm{AC}}$, blue arrow) encircle the signal line, enabling control of the spin states of the $\mathrm{V_{Si}}$ centers, whose spin quantization axes are perpendicular to the surface. Top views of (b) a schematic image with dimensions and (c) an optical microscopy image of the fabricated waveguide.
}
\end{figure}

We consider a quantum sensing system utilizing a coplanar waveguide-based microwave solution, as shown in Fig.~\ref{fig:fig1}(a). 
The system consists of a semiconductor substrate containing color centers and a signal line with a thin slit, positioned between the ground lines. 
When the slit width is sufficiently small, the microwave magnetic field distribution is almost unaffected by the introduction of the slit, resulting in the fields encircling the signal line [indicated by the blue arrow in Fig.~\ref{fig:fig1}(a)]. 
By irradiating an excitation laser onto color centers (green arrow) and collecting the photoluminescence (PL, red arrow) through this slit, optical measurements of color centers can be performed. 
Consequently, the slit-loaded coplanar waveguide enables the control of color center spins whose spin quantization axes are oriented perpendicular to the surface, such as the $\mathrm{V_{Si}}$ centers in 4H-SiC(0001).
This is because only in-plane microwave magnetic fields ($B_{\mathrm{AC},x}$ and $B_{\mathrm{AC},y}$ where the $z$-axis is normal to the surface) are capable of manipulating their spin states. 
Figure \ref{fig:fig1}(b) shows the schematic image of this waveguide with dimensions. 
The widths of the signal line and the gap between the signal and ground lines are $W_{\mathrm{signal}}=100\ \upmu$m and $W_{\mathrm{gap}}=40\ \upmu$m, respectively. 
The width of the ground lines, $W_{\mathrm{ground}}=200\ \upmu$m, is large enough to ensure the operation performance of the coplanar waveguide. 
Thus, the total waveguide width can be as small as 580 $\upmu$m. 
A thin slit with a width of $W_{\mathrm{slit}}=40\ \upmu$m and a length of $L_{\mathrm{slit}}=300\ \upmu$m is loaded within the signal line. 
In our experiments, the aluminum waveguide is fabricated on a 4H-SiC(0001) substrate with these parameter values, closely matching the design (within an error of 4 $\upmu$m), as shown in Fig \ref{fig:fig1}(c).

The dependence of the generated microwave magnetic fields on these design parameters of the slit-loaded coplanar waveguides ($W_{\mathrm{signal}}$, $W_{\mathrm{gap}}$, $W_{\mathrm{ground}}$, $W_{\mathrm{slit}}$, and $L_{\mathrm{signal}}$) is studied using the finite-integration-technique-based simulator CST Studio Suite$^{\text{\textregistered}}$ with an input power of 1 W. 
The waveguide has a characteristic impedance of 50 \textbf{$\Omega$} in order to efficiently apply microwaves while suppressing reflections at the input port. 
This condition is fulfilled within an error margin of ±3.5 \textbf{$\Omega$} for $W_{\mathrm{signal}}=100\ \upmu$m and $W_{\mathrm{gap}}=40\ \upmu$m, considering a SiC substrate thickness of 0.3 mm. 
The reflection is less than -30 dB over a wide frequency range of 70 MHz to 3 GHz, indicating efficient microwave irradiation even after loading the slit.
This frequency range covers the corresponding detection window of the longitudinal DC magnetic fields from 0 to over 1000 G using the $\mathrm{V_{Si}}$ centers in 4H-SiC.

\begin{figure}[!b]
\includegraphics{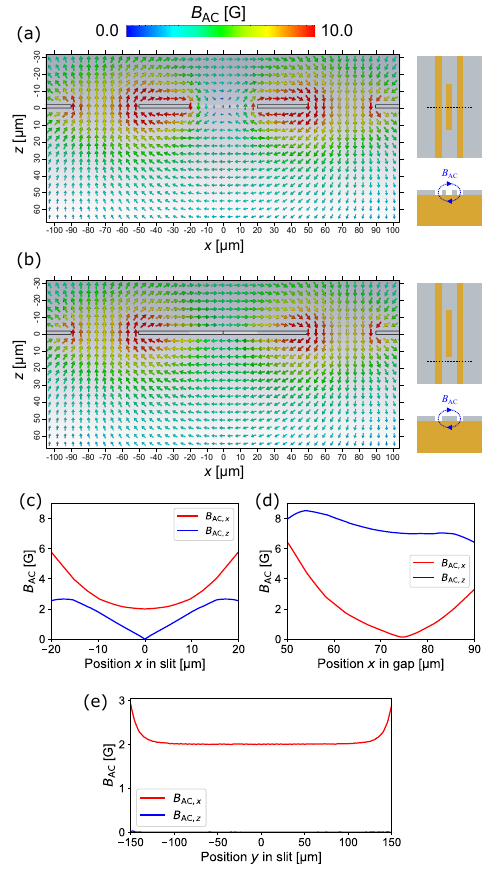}
\caption{\label{fig:fig2}
  (a, b) Cross-sectional views of the microwave magnetic field $\boldsymbol{B}_{\mathrm{AC}}$ distribution near the signal line. The direction and color of each arrow represent the direction and magnitude of $\boldsymbol{B}_{\mathrm{AC}}$, respectively. The black squares indicate conductors (signal and ground lines). Schematic images of the top and cross-sectional views are shown on the right side of the arrow plots. As shown by the dotted black lines in the top views, (a) and (b) are the results at the center of the slit ($y=0$) and in the area without the slit ($y=160\ \upmu$m), respectively. Note that the origin $(x, y, z) = (0, 0, 0)$ is located at the center of the slit at the surface of the SiC substrate. The dotted blue arrows represent the approximate direction of $\boldsymbol{B}_{\mathrm{AC}}$. (c-e) Cross-sectional line plots at a depth of $z=8.1 \ \upmu$m: (c, d) along the $x$-axis at $y=0$, showing the field distribution (c) within the slit and (d) within the gap between the signal line and the ground line; (e) along the $y$-axis at $x=0$ within the slit. $B_{\mathrm{AC},x}$ and $B_{\mathrm{AC},z}$ represent the $x$ and $z$ components of the microwave magnetic fields, respectively. All results are calculated at a microwave frequency of 70 MHz unless otherwise mentioned.
}
\end{figure}

Figure \ref{fig:fig2}(a) shows the simulated distribution of microwave magnetic fields in the sectional view of the signal line at the center of the slit with $W_{\mathrm{slit}}=40\ \upmu$m and $L_{\mathrm{slit}}=300\ \upmu$m. 
The origin of the Cartesian coordinate system, $(x, y, z) = (0, 0, 0)$, is located at the center of the slit on the surface of the SiC substrate. 
Thus, the slit area is $|x|\le20$ $\upmu$m and $|y|\le150$ $\upmu$m, with the $z$ value representing the depth from the surface. 
The two inner squares are the signal lines divided by the slit and the two outer squares are the ground lines placed on both sides of the signal line. 
The direction and color of each arrow represent the direction and magnitude of the microwave magnetic fields, respectively. 
To observe the effect of loading the slit, we also present the field distribution in the section without the loaded slit as a reference [Fig.~\ref{fig:fig2}(b)]. 
The position of each sectional view is shown in the corresponding schematic image on the right side of each arrow plot. 
Both the direction and the magnitude of the microwave magnetic fields remain mostly consistent, except in the area very close to the slit. 
This result indicates that the fundamental mode of a typical coplanar waveguide, where microwave magnetic fields encircle the signal line [blue dashed arrows in Figs. \ref{fig:fig2}(a, b)], is maintained even after loading the slit.
This is because the slit size is much smaller than the effective wavelength (i.e., several 10 mm, corresponding to several 10 MHz in the material). 
Therefore, the fields are approximately parallel to the SiC surface underneath the slit ($|x|\le20$ $\upmu$m) whereas their direction is nearly perpendicular to the surface in the gaps (50 $\upmu$m$\le\left|x\right|\le90$ $\upmu$m). 
It should be noted that although the simulation results for the microwave field distributions presented in this paper are calculated at a microwave frequency of 70 MHz, we have verified that the field distributions are almost independent of the frequency in the range of 70 MHz to 3 GHz.

These characteristics are clearly seen in the cross-sectional line plots [Figs. \ref{fig:fig2}(c, d)]. 
$B_{\mathrm{AC},x}$ and $B_{\mathrm{AC},z}$ represent the $x$ and $z$ components of the microwave magnetic fields, respectively. 
Given that the $y$ component $B_{\mathrm{AC},y}$ is negligibly small, the in-plane component can be considered equal to $B_{\mathrm{AC},x}$. 
This component is larger than the out-of-plane component $B_{\mathrm{AC},z}$ within the slit [Fig.~\ref{fig:fig2}(c)], whereas the opposite relation is observed within the gap [Fig.~\ref{fig:fig2}(d)]. 
Furthermore, $B_{\mathrm{AC},x}$ has larger magnitude within the slit than within the gap, excluding the very close region within $\sim$10 $\upmu$m of the signal line. 
The in-plane component $B_{\mathrm{AC},x}$ consistently exceeds 2 G within the slit; in contrast, it exhibits a drastic change and diminishes at a specific point in the gap. 
This result demonstrates the advantage of incorporating the slit to apply $B_{\mathrm{AC},x}$ more efficiently. 
The distribution of the microwave magnetic fields along the $y$-axis is shown in Fig.~\ref{fig:fig2}(e), indicating uniform $B_{\mathrm{AC},x}$ along the slit, which enables spin manipulation by $B_{\mathrm{AC},x}$ throughout the slit. 
The out-of-plane component $B_{\mathrm{AC},z}$ is constantly zero in Fig.~\ref{fig:fig2}(e) because we are observing a line right in the middle of the slit (i.e., at $x=0$).

\section{\label{sec:level1}Sample and experimental setup}

\begin{figure}[!b]
\includegraphics{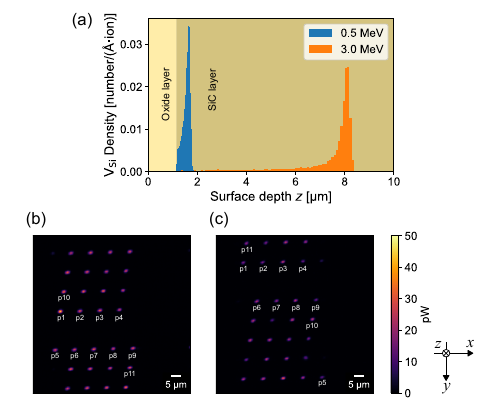}
\caption{\label{fig:fig3}
  (a) Simulated results showing the depth distribution of the created silicon vacancies in SiC. The peak depths, where the vacancy distribution is maximized, are estimated to be approximately 1.7 $\upmu$m and 8.1 $\upmu$m for the 0.5 MeV and 3.0 MeV implanted samples, respectively. Note that the sample has a 1.15 $\upmu$m-thick oxide layer on its surface. (b, c) CFM scanning images with an area of $80\times80\ \upmu\text{m}^2$. The energy of the He ion beam is set to (b) 0.5 MeV and (c) 3.0 MeV. The measurement spots are labeled as p1, p2, … p11 in each sample.
}
\end{figure}

Room temperature measurements are performed using a home-built confocal fluorescence microscope (CFM).
A pulse-controllable 785 nm excitation laser (Omicron LuxX+ 785-200) is passed through a lens-pinhole-lens spatial filter and directed onto the sample via an 850 nm cut-on dichroic mirror and an objective lens (Olympus LCPLN50XIR) mounted on a three-axes piezostage. 
The laser power is 5 mW measured in front of the objective. 
The PL from the sample is collected via the objective lens and passes through a dielectric mirror, a 900 nm long-pass filter, and a 785 nm notch filter. 
A pinhole aperture is placed to extract the focused PL, which is converted into an analog voltage signal by a photoreceiver (FEMTO OE-200-IN1, gain set at $10^{11}$ V/W) and then recorded using a data acquisition device (National Instruments USB-6363). 
DC magnetic fields from a permanent SmCo magnet are applied to the sample. 
To conduct ODMR and Rabi measurements, a microwave signal from a signal generator (Anritsu MG3710E) is pulse-controlled using a microwave switch (Mini-Circuits ZYSWA-2-50DR+) and a data-timing generator (DTG5274), and then amplified by a Mini-Circuits ZHL-20W-202-S+ amplifier. 
The magnitudes of both forward and reflected microwaves are monitored by high-frequency power meters (R\&S NRP-Z22). Our sample is mounted on a printed circuit board with inter-pad connections formed by wire bonding. 
To control these measurement instruments with synchronized timing and display results in real-time, we use MAHOS software \cite{ref41}.

We use a $9 \times 9 \times 0.3 \ \ \text{mm}^3$ $n$-type 4H-SiC(0001) single crystal substrate. 
Aluminum waveguide patterns with a thickness of 2 $\upmu$m are deposited onto the SiC substrate via sputtering and photolithography. 
To create the $\mathrm{V_{Si}}$ centers within the slit, particle beam writing (PBW) using a focused helium ion beam is performed on the patterned SiC sample \cite{ref42,ref43} at the TIARA irradiation facilities of QST Takasaki, Japan. 
Irradiation energies of 0.5 and 3.0 MeV are employed, corresponding to estimated surface depths of approximately 1.7 and 8.1 $\upmu$m, respectively, as obtained by Monte-Carlo simulations using the well-established SRIM software \cite{ref44} [Fig.~\ref{fig:fig3}(a)]. 
Although the SiC sample has a 1.15 $\upmu$m oxide layer on its surface, the implantation energy of PBW is sufficient for ion beams to penetrate the topmost oxide layer and reach the SiC substrate. 
We consider that the oxide layer has little effect on the microwave magnetic field distributions. 
Using the PBW technique, uniformly spaced, dense $\mathrm{V_{Si}}$ ensemble clusters with a $\sim 0.5 \ \text{$\upmu$m}^2$ spot area are formed on a rectangular grid with $\sim$20 $\upmu$m spacing. 
The number of ions implanted in each spot is $3 \times 10^5$. The formation of these clusters is confirmed by CFM images, as shown in Figs. \ref{fig:fig3}(b, c). 
Typical PL intensities of each spot are about 50$\sim$60 pW in the 0.5 MeV-implanted sample [Fig.~\ref{fig:fig3}(b)], which is higher than that of the 3.0 MeV-implanted sample (20$\sim$30 pW) [Fig.~\ref{fig:fig3}(c)] due to the higher extraction efficiency of PL from the shallower $\mathrm{V_{Si}}$ centers. 
The patterned $\mathrm{V_{Si}}$ clusters created with different irradiation energies enable the measurement of the spatial distribution of the microwave magnetic fields in the slit.

\section{\label{sec:level1}Experimental results}

\begin{figure}[!b]
\includegraphics{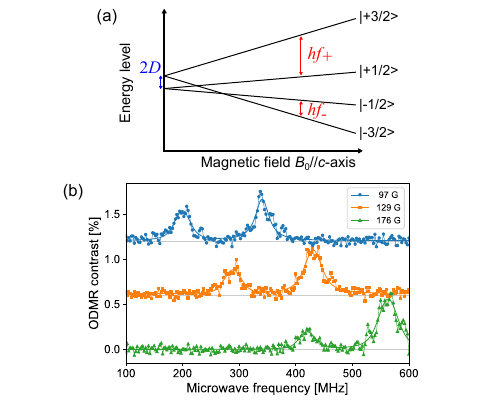}
\caption{\label{fig:fig4}
  (a) Schematic image of the spin ground states of the $\mathrm{V_{Si}}$ centers under external static magnetic fields along the c-axis of the SiC(0001) crystal. (b) Examples of CW-ODMR measurement results at point p1 in the 3.0 MeV-implanted sample, as shown in Fig \ref{fig:fig3}(c). The applied static magnetic fields along the $c$-axis ($B_0$) are 97 G (blue), 129 G (orange) and 176 G (green). To improve visibility, an offset has been added in the vertical direction and the baselines for each dataset are plotted as gray lines. For each $B_0$ condition, the lower (higher) ODMR frequency corresponds to $f_-$ ($f_+$). The solid lines denote the fitting results using the double Voigt peak model.
}
\end{figure}

To demonstrate the control of the $\mathrm{V_{Si}}$ spins using the microwave magnetic fields generated from the slit-loaded waveguide, we perform CW-ODMR measurements. 
The spin Hamiltonian $\mathcal{H}$ for the $\mathrm{V_{Si}}$ centers with $S=3/2$ is described as:

\begin{equation}
    \mathcal{H}=D\left[S_z^2-S\left(S+1\right)/3\right]+g_e\mu_B\boldsymbol{B}_0\cdot\boldsymbol{S}, \label{eq:hamiltonian}
\end{equation}

\noindent
where $g_e\approx2.0$ is the electron g-factor, $\mu_{\mathrm{B}}$ is the Bohr magneton, $S_z$ is the projection of the total spin $S=3/2$ on the c-axis of SiC, $\boldsymbol{B}_0$ is the external static magnetic field, and $D$ is the zero-field-splitting parameter \cite{ref20,ref21,ref22}. 
Here, the hyperfine interaction and the transverse zero-field splitting parameter $E$ are neglected in Eq.~\eqref{eq:hamiltonian}. 
At zero field ($B_0=0$, where $B_0=|\boldsymbol{B}_0|$), the ground state of the $\mathrm{V_{Si}}$ electron spin exhibits a zero-field-splitting of $2D/h\approx70$ MHz between two spin levels, $m_s = \pm 3/2$ and $m_s = \pm 1/2$, where $h$ is Planck’s constant. 
By applying $\boldsymbol{B}_0$ along the $z$-axis (i.e., parallel to the c-axis of SiC), the two resonance frequencies, 

\begin{equation}
    f_+=2D/h+g\mu_BB_0/h \label{eq:f_plus},
\end{equation}

\noindent
and

\begin{equation}
    f_-=\left|2D/h-g\mu_BB_0/h\right| \label{eq:f_minus},
\end{equation}

\noindent
corresponding to the two allowed transitions $\left|\left.+3/2\right\rangle\leftrightarrow\left|\left.+1/2\right\rangle\right.\right.$ and $\left|\left.-3/2\right\rangle\leftrightarrow\left|\left.-1/2\right\rangle\right.\right.$, respectively, are mainly observed in CW-ODMR measurements of $\mathrm{V_{Si}}$ [Fig.~\ref{fig:fig4}(a)]. 
The ODMR signals appear as peaks because the PL intensity is higher when the spin is excited from $\left|\left.\pm3/2\right\rangle\right.$ than from $\left|\left.\pm1/2\right\rangle\right.$ and laser irradiation initializes the spin state into $\left|\left.\pm1/2\right\rangle\right.$. 

Figure \ref{fig:fig4}(b) shows an example of the measured ODMR spectra at point p1 in the 3.0 MeV-implanted sample under three different $B_0$ conditions. 
We define the ODMR contrast as $\left(I_1-I_0\right)/I_0$, where $I_0$ and $I_1$ are the measured PL intensity without and with a microwave input, respectively, as the microwave frequency is swept. 
In each dataset, the ODMR contrast is about 0.3$\sim$0.5\%. 
The double Voigt peak model is in good agreement with the measured data and the applied $B_0$ is estimated to be 97 G, 129 G and 176 G for the blue, orange, and green ODMR spectra, respectively. 
These results confirm that our waveguide generates sufficient $B_{\mathrm{AC},x}$ to manipulate the color center spins while allowing both the irradiation of the excitation laser and the extraction of PL from the color centers through the slit.

\begin{figure}[!b]
\includegraphics{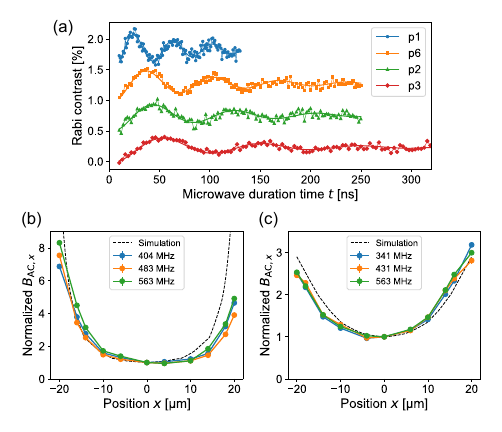}
\caption{\label{fig:fig5}
  (a) Example Rabi oscillation measurements at points p1, p2, p3, and p6 in the 3.0 MeV-implanted sample at 97 G. The solid lines show the fitting results using Eq.~\eqref{eq:Rabi}. Offsets have been added along the vertical axis for clarity. (b, c) Magnitude of the microwave magnetic fields parallel to the SiC surface, obtained from the measured Rabi frequencies. The ion implantation energy is (b) 0.5 MeV (depth 1.7 $\upmu$m) and (c) 3.0 MeV (depth 8.1 $\upmu$m). For (b), resonant microwave frequencies are 404 MHz (blue), 483 MHz (orange), and 563 MHz (green) under the static magnetic fields of 111 G, 148 G, and 176 G, respectively; for (c), they are 341 MHz (blue), 431 MHz (orange), and 563 MHz (green) under the static magnetic fields of 97 G, 129 G, and 176 G, respectively. For comparison, the simulation results are plotted as black dashed lines. Datasets are normalized relative to $B_{\mathrm{AC},x}$ at $x=0$.
}
\end{figure}

To quantitatively evaluate $B_{\mathrm{AC},x}$, Rabi oscillation measurements are also conducted. 
Following a 0.5 $\upmu$s laser pulse for initializing the spin state into $\left|\left.\pm1/2\right\rangle\right.$, a microwave pulse with a frequency of $f_\pm$ and a duration of $t$ is applied. 
The PL intensity is then read out using a final laser pulse. 
A delay of 0.7 $\upmu$s is introduced between the initialization and the microwave pulse application to ensure complete decay from the metastable doublet states to the ground state. 
To observe Rabi oscillations with a slow (1.1 kHz) photoreceiver, we measure the average PL intensity during repeated pulse sequences within fixed (1/60 s) time windows \cite{ref45}. 
During measurements, the output power of the signal generator is kept constant. 
The Rabi contrast is defined as $C\left(t\right)=\left(I_1-I_0\right)/I_0$. 
Figure \ref{fig:fig5}(a) shows examples of the measured Rabi oscillation data (circles) at point p1 ($x=-20\ \upmu$m, the edge of the slit), p6 ($x=-16\ \upmu$m), p2 ($x=-10\ \upmu$m) and p3 ($x=0\ \upmu$m, the center of the slit) in the 3.0 MeV-implanted sample. 
For all data, the resonant microwave frequency is $f_+=340.76$ MHz under $B_0=97$ G. 
The fitting results (solid lines) are also plotted using the following model:

\begin{equation}
    C\left(t\right)=-A_{\mathrm{Rabi}}\cos{\left(2\pi f_{\mathrm{Rabi}}t\right)e^{-t/T_2^\ast}} \label{eq:Rabi},
\end{equation}

\noindent
where $A_{\mathrm{Rabi}}$ and $f_{\mathrm{Rabi}}$ are Rabi contrast and frequency, respectively. 
As shown in Fig.~\ref{fig:fig5}(a), $f_{\mathrm{Rabi}}$ varies with the position $x$ in the slit; the closer to the slit edge, the faster the Rabi oscillation becomes, indicating larger $B_{\mathrm{AC},x}$.

To convert $f_{\mathrm{Rabi}}$ to in-plane microwave magnetic fields $B_{\mathrm{AC},x}$, the following formula for the $S=3/2$ spin system is used \cite{ref22}:

\begin{equation}
    f_{\mathrm{Rabi}} = \sqrt{3} g \mu_{\mathrm{B}} B_{\mathrm{AC},x} / h \label{eq:Rabi2power}.
\end{equation}

\noindent
The obtained $B_{\mathrm{AC},x}$ values are co-plotted in Figs. \ref{fig:fig5}(b, c) with the simulation results as a function of position x. 
Each dataset is normalized by $B_{\mathrm{AC},x}$ value at $x=0$. 
The behaviors of the measured $B_{\mathrm{AC},x}$ as a function of $x$ are in good agreement with the simulation results at both surface depths of 1.7 and 8.1 $\upmu$m, validating the simulated microwave magnetic field distributions within the slit. 
The slight mismatch of $B_{\mathrm{AC},x}$ as a function of $x$ can be explained by the slit width $W_{\mathrm{slit}}$, as the $W_{\mathrm{slit}}$ of the fabricated waveguide is about 4 $\upmu$m wider than the design value. 
Note that we normalize the plotted data to compare them since the microwave power applied to the waveguide cannot be precisely estimated due to the microwave reflection mainly at the thin bonding wires.

\section{\label{sec:level1}Discussion}

We have demonstrated that by loading a slit in the signal line of the coplanar waveguide, the generated in-plane microwave magnetic fields $B_{\mathrm{AC},x}$ can efficiently manipulate the spin states of the $\mathrm{V_{Si}}$ centers, while the excitation laser and the emitted PL from the color centers can pass through the slit. 
As explained in Section II, the loaded slit hardly affects the microwave magnetic field distribution of the fundamental mode of a typical coplanar waveguide as shown in Figs. \ref{fig:fig2}(a, b), which results in a $B_{\mathrm{AC},x}$ of sufficient magnitude in the slit. 
Compared with microwave antennas that can generate $B_{\mathrm{AC},x}$ \cite{ref27}, a key advantage of our waveguide-based system is its capability to operate over a broad frequency range and its facilitation of a compact microwave system not restricted by the microwave wavelength.
It has been reported that a multi-strip-line waveguide can apply a spatially homogeneous $B_{\mathrm{AC},x}$, which requires a ground plane on the backside of the device \cite{ref46}. 
By contrast, the configuration of our slit-loaded coplanar waveguide offers the advantage that both the signal and ground lines of the waveguide are in the same plane, which simplifies fabrication.

\begin{figure}[!b]
\includegraphics{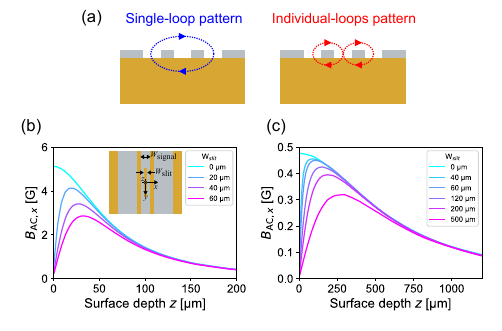}
\caption{\label{fig:fig6}
  (a) Illustration of the two types of loop patterns for the microwave magnetic field distribution [Fig.~\ref{fig:fig2}(a)]. The left illustration represents the single-loop pattern (blue dotted line), and the right illustration represents the individual-loops pattern (red dotted lines). (b, c) Dependence of the in-plane microwave magnetic fields $B_{\mathrm{AC},x}$ on the depth $z$ from the surface of the SiC substrate for various $W_{\mathrm{slit}}$. The vertical axis denotes the microwave magnetic fields parallel to the surface. The signal line width $W_{\mathrm{signal}}$ is set to (b) 100 $\upmu$m and (c) 1 mm. The inset of (b) is the top view of the schematic image of the waveguide. All results are calculated at 70 MHz.
}
\end{figure}

We further discuss the depth dependence of $B_{\mathrm{AC},x}$ generated by the slit-loaded coplanar waveguide. 
Focusing on the microwave magnetic field distribution in the slit near the SiC surface (i.e., at $z\approx0$), a reduction of the magnitude of $\boldsymbol{B}_{\mathrm{AC}}$ can be observed [Fig.~\ref{fig:fig2}(a)]. 
This reduction arises from the characteristic microwave magnetic field distribution, as shown in the left illustration of Fig.~\ref{fig:fig6}(a), where the fields encircle each of the split signal lines in the same direction of rotation. 
At the origin, the field components encircling the left and right split signal lines are antiparallel, leading to completely canceled fields (i.e., $B_{\mathrm{AC},x}=0$), as shown in Fig.~\ref{fig:fig2}(a) at $\left(x,z\right)=\left(0,0\right)$. 
As the depth increases, moving away from the slit, the dominant contribution gradually changes from the individual-loops pattern to the single-loop pattern (see Fig.~\ref{fig:fig6}(a), left and right illustrations, respectively). 
Note that the single-loop pattern around the entire signal line arrangement indicates the fundamental mode of the coplanar waveguide, as previously mentioned. 
This behavior results in the enhancement of $B_{\mathrm{AC},x}$ at larger $z$ values. 
Meanwhile, as a general characteristic, the magnitude of the microwave magnetic fields decays with increasing distance from the signal line. 
Consequently, $B_{\mathrm{AC},x}$ reaches its maximum value at a certain depth.

These characteristics of the depth dependence of $B_{\mathrm{AC},x}$ are summarized in Fig.~\ref{fig:fig6} for various slit widths $W_{\mathrm{slit}}$ and signal line widths $W_{\mathrm{signal}}$. Note that the in-plane position is fixed at the center of the slit $(x,y)=(0,0)$. 
In Fig.~\ref{fig:fig6}(a), the purple line represents the result for $W_{\mathrm{slit}}=40\ \upmu$m and $W_{\mathrm{signal}}=100\ \upmu$m, which is the same as the waveguide used in our measurements, and shows a maximum $B_{\mathrm{AC},x}$ value of 3.4 G at $z=26\ \upmu$m. 
These results suggest that larger $B_{\mathrm{AC},x}$ can be applied if the device structure is optimized to form color centers at greater depths than those of the samples used in our measurements. 
Because the individual-loops pattern does not appear in a normal coplanar waveguide without a slit, $B_{\mathrm{AC},x}$ monotonically decreases with depth, as shown by the cyan line in Fig.~\ref{fig:fig6}(a). 
The simulation results show that as $W_{\mathrm{slit}}$ increases, the maximum $B_{\mathrm{AC},x}$ value decreases and shifts to greater depths, indicating a larger contribution of the individual-loops pattern. 
We also conduct simulations with a wider signal line of $W_{\mathrm{signal}}=1$ mm, as shown in Fig.~\ref{fig:fig6}(b). 
Compared with the case of $W_{\mathrm{signal}}=100\ \upmu$m, $B_{\mathrm{AC},x}$ is an order of magnitude smaller. 
This is attributable to the less concentrated electromagnetic fields in the wider signal line, resulting in weaker local microwave magnetic fields. 
Consequently, the benefit of downsizing the waveguide is twofold: it facilitates the fabrication of color-center-based sensing systems in semiconductor devices and enhances the efficiency of microwave generation. 
In addition, $B_{\mathrm{AC},x}$ reaches its maximum value at a much deeper position in the wider signal line. 
This behavior can also be attributed to the individual-loops-pattern's contribution, which is governed by the ratio of $W_{\mathrm{slit}}$ (i.e., the distance between the split signal lines) to $W_{\mathrm{signal}}$.

\section{\label{sec:level1}Conclusions}

This study has presented a slit-loaded coplanar waveguide for realizing a compact microwave system for controlling color center spins. 
The loaded slit in the signal line, while allowing the transmission of the excitation laser and photoluminescence from color centers, hardly affects the microwave magnetic field distribution around the signal line, thereby providing in-plane microwave magnetic fields. 
These in-plane fields are crucial for manipulating the spin states of color centers with out-of-plane spin quantization axes, such as the $\mathrm{V_{Si}}$ (V2) centers in 4H-SiC(0001) and NV centers in diamond(111). 
CW-ODMR and Rabi oscillation measurements using the designed waveguide demonstrated quantitative agreement with simulation results. 
Our waveguide-based system operates over a wide microwave frequency range (tens of MHz to GHz) and can be integrated with a width as small as several hundred micrometers. 
This compact design offers an efficient way of microwave generation due to the concentrated electromagnetic fields in the narrow signal line, making it suitable for applications in semiconductor electronic devices and other integrated systems.

\begin{acknowledgments}
This work was supported by the Council for Science, Technology and Innovation (CSTI), Cross-ministerial Strategic Innovation Promotion Program (SIP), under the research theme “Environment development for practical use of solid-state quantum sensors: towards social implementation,” within the project “Promoting application of advanced quantum technologies to social challenges” (Funding agency: QST).

The authors also thank Hidekazu Tsuchida at the Central Research Institute of Electric Power Industry for the SiC epitaxial growth.

\end{acknowledgments}

\bibliography{reference}

\end{document}